\documentclass[11pt,a4paper,final]{article}
\usepackage{amsmath}%
\usepackage{amstext}%
\usepackage{amssymb}%
\usepackage{showkeys}%
\usepackage{epsfig}%
\usepackage{graphicx}%
\usepackage{xcolor}
\usepackage{multicol}
\usepackage{amsthm}
\usepackage{booktabs}
\usepackage{apacite}
\usepackage{multirow}
\usepackage[top=1in, bottom=1.25in, left=0.8in, right=0.8in]{geometry}

\theoremstyle{plain}
\theoremstyle{proof}

\newtheorem{proposition}{Proposition}
\newtheorem{algorithm}{Algorithm}
\newtheorem{assumption}{Assumption}
\newtheorem{theorem}{Theorem}
\newtheorem{lemma}[theorem]{Lemma}

\providecommand{\keywords}[1]{
\textbf{Keywords:~~~} Bayesian MS--VAR process, Stochastic DDM, Gordon's Growth Model.
}

\begin{document}
\title{EM Estimation of Conditional Matrix Variate $t$ Distributions}
\author{Battulga Gankhuu\footnote{Department of Applied Mathematics, National University of Mongolia; E-mail: battulgag@num.edu.mn; Phone Number: 976--99246036}}
\date{}

\maketitle 

\begin{abstract}
Conditional matrix variate student $t$ distribution was introduced by \citeA{Battulga24g}. In this paper, we propose a new version of the conditional matrix variate student $t$ distribution. The paper provides EM algorithms, which estimate parameters of the conditional matrix variate student $t$ distributions, including general cases and special cases with Minnesota prior.
\end{abstract}


\section{Introduction}

An expectation--maximization (EM) algorithm was proposed and given its name by \citeA{Dempster77}. The EM algorithm is an iterative method to obtain (local) maximum likelihood estimates of parameters of distribution functions, which depend on unobserved (latent) variables. The EM algorithm alternates an expectation (E) step and a maximization (M) step. In the E--step, one considers that conditional on available data and the current estimate of the parameters, expectation of augmented log--likelihood of the data, and unobserved (latent) variables. The E--Step defines an objective function. In the M--step, to obtain a parameter estimate of the next iteration, one maximizes the objective function with respect to the parameters. Alternating between these steps, the EM algorithm produces improved parameter estimates at each step (in the sense that the value of the original log--likelihood is continually increased), and it converges to the maximum likelihood (ML) estimates of the parameters.

The EM algorithm is widely used in econometrics. In particular, \citeA{Hamilton90} introduced a parameter estimation method for a general regime-switching model. The regime-switching model assumes that a discrete unobservable Markov process randomly switches among a finite set of regimes and that a particular parameter set defines each regime.
In finance, to value private companies whose market prices are unobservable, \citeA{Battulga23b} and \citeA{Battulga24c} applied the EM algorithm. \citeA{McNeil05} provides an EM algorithm to estimate parameters of the generalized hyperbolic distribution, which can be used to model financial returns. Also, the EM algorithm has been used in classifications. For example, to estimate matrix variate $t$ distribution parameters, \citeA{Thompson20} used the EM algorithm. \citeA{Sun10} provided an EM algorithm to estimate the parameters of Pearson VII distribution.

Classic Vector Autoregressive (VAR) process was proposed by \citeA{Sims80} who criticize large--scale macro--econometric models, which are designed to model interdependencies of economic variables. Besides \citeA{Sims80}, there are some other important works on multiple time series modeling, see, e.g., \citeA{Tiao81}, where a class of vector autoregressive moving average models was studied. For the VAR process, a variable in the process is modeled by its past values and the past values of other variables in the process. After the work of \citeA{Sims80}, VARs have been used for macroeconomic forecasting and policy analysis. However, if the number of variables in the system increases or the time lag is chosen high, then too many parameters need to be estimated. This will reduce the degrees of freedom of the model and entail a risk of over--parametrization. 

Therefore, to reduce the number of parameters in a high--dimensional VAR process, \citeA{Litterman79} introduced probability distributions for coefficients that are centered at the desired restrictions but that have a small and nonzero variance. Those probability distributions are known as Minnesota prior in Bayesian VAR (BVAR) literature, which is widely used in practice. Due to over--parametrization, the generally accepted result is that the forecast of the BVAR model is better than the VAR model estimated by the frequentist technique. The BVAR relies on Monte--Carlo simulation. Recently, for Bayesian Markov--Switching VAR process, \citeA{Battulga24g} introduced a new Monte--Carlo simulation method that removes duplication in a regime vector. Also, the author introduced importance sampling method to estimate probability of rare event, which corresponds to endogenous variables. Research works have shown that BVAR is an appropriate tool for modeling large data sets; for example, see \citeA{Banbura10}. 

The rest of the paper is organized as follows: In Section 2, for type I conditional matrix variate distributions, including general case and special case with Minnesota prior we develop EM algorithms. Section 3 is dedicated to studying EM algorithms for type II conditional matrix variate distributions, including general case and special case with Minnesota prior. Finally, Section 4 concludes the study.

\section{EM Estimation of Type I Conditional Matrix Variate $t$ Distribution}

We consider a Bayesian Vector Autoregressive process of $p$ order (BVAR($p$)), which is given by the following equation 
\begin{equation}\label{08001}
y_t=A_{0,t}\psi_t+A_{1,t}y_{t-1}+\dots +A_{p,t}y_{t-p}+\xi_t,~t=1,\dots,T,
\end{equation}
where $y_t=(y_{1,t},\dots,y_{n,t})'$ is an $(n\times 1)$ vector of endogenous variables, $\psi_t=(1,\psi_{2,t},\dots,\psi_{l,t})'$ is an $(l\times 1)$ vector of exogenous variables, $\xi_t=(\xi_{1,t},\dots,\xi_{n,t})'$ is an $(n\times 1)$ residual process, $A_{0,t}$ is an $(n\times l)$ random coefficient matrix, corresponding to the vector of exogenous variables, for $i=1,\dots,p$, $A_{i,t}$ are $(n\times n)$ random coefficient matrices, corresponding to $y_{t-1},\dots,y_{t-p}$. Equation \eqref{08001} can be written by
\begin{equation}\label{08002}
y_t=\Pi_t\mathsf{Y}_t+\xi_t,~t=1,\dots,T,
\end{equation}
where $\Pi_t:=[A_{0,t}: A_{1,t}:\dots:A_{p,t}]$ is an $(n\times d)$ random coefficient matrix with $d:=l+np$, which consist of all the random coefficient matrices and $\mathsf{Y}_t:=(\psi_t',y_{t-1}',\dots,y_{t-p}')'$ is a $(d\times 1)$ vector, which consist of exogenous variable $\psi_t$ and last $p$ lagged values of the process $y_t$. The process $\mathsf{Y}_t$ is measurable with respect to a $\sigma$--field $\mathcal{F}_{t-1}$, which is defined below. Let us collect endogenous variables into ($[nT]\times 1$) vector $y$, i.e., $y:=(y_1',\dots,y_T')'$.

For the residual process $\xi_t$, we assume that it has $\xi_t:=\Sigma_t^{1/2}\varepsilon_t$, $t=1,\dots,T$ representation, where $\Sigma_t^{1/2}$ is a Cholesky factor of a positive definite $(n\times n)$ random matrix $\Sigma_t$ and $\varepsilon_1,\dots,\varepsilon_T$ is a random sequence of independent identically multivariate normally distributed random vectors with means of 0 and covariance matrices of $n$ dimensional identity matrix $I_n$. We also assume that the strong white noise process $\{\varepsilon_t\}_{t=1}^T$ is independent of the random coefficient matrices $(\Pi_1,\dots,\Pi_T)$ and $(\Sigma_1,\dots,\Sigma_T)$ conditional on initial information $\mathcal{F}_0:=\{y_{1-p},\dots,y_0,\psi_{1},\dots,\psi_T\}$, where $\psi_1,\dots,\psi_T$ are values of exogenous variables and they are known at time zero. We also denote available information at time $t$ by $\mathcal{F}_t:=\{\mathcal{F}_0,y_1,\dots,y_t\}$.

\subsection{General Type I Conditional Matrix Variate $t$ Distribution}

Let us assume that for $t=1,\dots,T$, the random coefficient matrices $\Pi_t$ and random covariance matrices $\Sigma_t$ are equals, that is, $\Pi:=\Pi_1=\dots=\Pi_T$ and $\Sigma:=\Sigma_1=\dots=\Sigma_T$. Then,
by using the Kronecker product, the BVAR($p$) process can be written by the following equation  
\begin{equation}\label{08003}
y_t=\Pi\mathsf{Y}_{t}+\xi_t=(\mathsf{Y}_{t}'\otimes I_n)\pi+\xi_t, ~~~t=1,\dots,T,
\end{equation}
where $\otimes$ is the Kronecker product of two matrices and $\pi:=\text{vec}(\Pi)$ is an $(nd\times 1)$ vectorization of the random coefficient matrix $\Pi$. Now we define distributions of the random coefficient vector $\pi$ and covariance matrix $\Sigma$. We assume that conditional on the initial information $\mathcal{F}_0$, a distribution of the random covariance matrix $\Sigma$ is given by 
\begin{equation}\label{08004}
\Sigma~|~\mathcal{F}_0\sim \mathcal{IW}(\nu_0,V_0)
\end{equation}
where the notation $\mathcal{IW}$ denotes the Inverse--Wishart distribution, $\nu_0>n-1$ is a degrees of freedom and $V_0$ is a positive definite scale matrix. Consequently, a distribution of the residual vector $\xi_t$ equals 
\begin{equation}\label{08005}
\xi_t~|~\Sigma,\mathcal{F}_0\sim \mathcal{N}\big(0,\Sigma\big),
\end{equation}
where $\mathcal{N}$ denotes the normal distribution. Also, we assume that conditional on the covariance matrix $\Sigma$ and initial information $\mathcal{F}_0$, a distribution of the random coefficient vector $\pi$ is given by 
\begin{equation}\label{08006}
\pi~|~\Sigma,\mathcal{F}_0\sim \mathcal{N}\Big(\pi_0,\Lambda_0\otimes \Sigma\Big),
\end{equation}
where $\pi_0$ is an $(nd\times 1)$ vector and $\Lambda_{0}=\text{diag}\{\lambda_{1},\dots,\lambda_{d}\}$ is a diagonal $(d\times d)$ matrix. Then, according to \citeA{Battulga24g}, the following Proposition holds.

\begin{proposition}\label{prop01}
Let $\pi~|~\Sigma,\mathcal{F}_0\sim \mathcal{N}\big(\pi_0,\Lambda_0\otimes \Sigma\big)$, and $\Sigma~|~\mathcal{F}_0\sim \mathcal{IW}(\nu_0,V_0)$. Then, first, conditional on the initial information $\mathcal{F}_0$, a joint density function of the random vector $\bar{y}_t$ is given by
\begin{eqnarray}\label{08007}
f(y|\mathcal{F}_0)=\frac{1}{\pi^{nT/2}}\frac{|\Lambda_0^{-1}|^{n/2}\Gamma_{n}\big((\nu_0+T)/2\big)|V_0|^{\nu_0/2}}{|\Lambda_{0|T}|^{n/2}\Gamma_{n}(\nu_0/2)\big|B_T+V_0\big|^{(\nu_0+T)/2}},
\end{eqnarray}
where $\Gamma_n(\cdot)$ is the multivariate gamma function, $\Lambda_{0|T}^{-1}:=\mathsf{Y}^\circ(\mathsf{Y}^\circ)'+\Lambda_0^{-1}$ is a $(d\times d)$ matrix, $y^\circ:=[y_1:\dots:y_T]$ is an $(n\times T)$ matrix, $\mathsf{Y}^\circ:=[\mathsf{Y}_0:\dots:\mathsf{Y}_{T-1}]$ is a $(d\times T)$ matrix, $\pi_0^\circ:=\mathbb{E}\big(\Pi\big|\Sigma,\mathcal{F}_0\big)$ is an $(n\times d)$ matrix, and $B_T$ is an  $(n\times n)$ positive semi--definite matrix and equals 
\begin{eqnarray}\label{08008}
B_T:=\big(y^{\circ}-\pi_0^{\circ}\mathsf{Y}^\circ\big)\big(I_T+(\mathsf{Y}^\circ)'\Lambda_0\mathsf{Y}^\circ\big)^{-1}\big(y^{\circ}-\pi_0^{\circ}\mathsf{Y}^\circ\big)'.
\end{eqnarray}
Second, conditional on the random covariance matrix $\Sigma$ and information $\mathcal{F}_t$, a joint density function of the random coefficient vector $\pi$ is given by
\begin{eqnarray}\label{08009}
f(\pi|\Sigma,\mathcal{F}_T)= \frac{1}{(2\pi)^{nd/2}|\Lambda_{0|T}|^{n/2} |\Sigma|^{d/2}} \exp\bigg\{-\frac{1}{2}\Big(\pi-\pi_{0|T}\Big)'\big(\Lambda_{0|T}^{-1}\otimes \Sigma^{-1}\big)\Big(\pi-\pi_{0|T}\Big)\bigg\},
\end{eqnarray}
where $\pi_{0|T}:=\big((\Lambda_{0|T}\mathsf{Y}^\circ)\otimes I_n\big)y+\big((\Lambda_{0|T}\Lambda_0^{-1})\otimes I_n\big)\pi_0$ is an $([nd]\times 1)$ vector. Finally, conditional on the information $\mathcal{F}_T$, a joint density function of the random coefficient matrix $\Sigma$ is given by
\begin{eqnarray}\label{08010}
f(\Sigma|\mathcal{F}_T)&=&\frac{\big|B_T+V_0\big|^{(\nu_0+T)/2}}{\Gamma_n\big((\nu_0+T)/2\big)2^{n(\nu_0+T)/2}}|\Sigma|^{-(\nu_0+T+n+1)/2}\exp\bigg\{-\frac{1}{2}\mathrm{tr}\Big(\big(B_T+V_0\big)\Sigma^{-1}\Big)\bigg\}.
\end{eqnarray}
\end{proposition}

The joint density function \eqref{08007} is called conditional matrix variate student $t$ density, see \citeA{Battulga24g}. To differentiate from a conditional matrix $t$ distribution, which arise in the following section, we refer to the distribution as type I conditional matrix variate student $t$ distribution. From equations \eqref{08009} and \eqref{08010}, one recognizes that posterior distributions of the random coefficient vector $\pi$ and covariance matrix $\Sigma$ are multivariate normal $\mathcal{N}\big(\pi_{0|t},\Lambda_{0|T}\otimes \Sigma\big)$ and inverse--Wishart $\mathcal{IW}\big(\nu_0+T,B_T+V_0\big)$, respectively. Let us denote a vector of all the parameters of the joint density function by $\theta:=\text{vec}(\pi_0,\Lambda_0,\nu_0,V_0)$. In this section, we develop Expectation--Maximization (EM) algorithm to estimate parameters of the density function. In E--Step, we consider that conditional on the full information $\mathcal{F}_T$ and parameter at iteration $k$, $\theta^{[k]}$, expectation of augmented log--likelihood of the data $y$ and unobserved (latent) variables $\pi$ and $\Sigma$. The E--Step defines a objective function $\mathcal{L}$, namely, 

\begin{eqnarray}\label{08011}
\mathcal{L}&=&\mathbb{E}\bigg[-\frac{n(T+d)}{2}\ln(2\pi)+\frac{T+d+\nu_0+n+1}{2}\ln|\Sigma^{-1}|\nonumber\\
&-&\frac{1}{2}\sum_{t=1}^T\big(y_t-(\mathsf{Y}_t'\otimes I_n)\pi\big)'\Sigma^{-1}\big(y_t-(\mathsf{Y}_t'\otimes I_n)\pi\big)\nonumber\\
&+&\frac{n}{2}\ln|\Lambda_0^{-1}|-\frac{1}{2}(\pi-\pi_0)'(\Lambda_0^{-1}\otimes \Sigma^{-1})(\pi-\pi_0)\\
&+&\frac{\nu_0}{2}\ln|V_0|-\ln\bigg(\Gamma_n\bigg(\frac{\nu_0}{2}\bigg)\bigg)-\frac{n\nu_0}{2}\ln(2)-\frac{1}{2}\mathrm{tr}\big(V_0\Sigma^{-1}\big)\bigg|\mathcal{F}_T;\theta^{[k]}\bigg]\nonumber
\end{eqnarray}

In M--Step, to obtain parameter estimate of next iteration $\theta^{[k+1]}$, one maximizes the objective function with respect to the parameter $\theta$. First, let us consider partial derivative from the objective function with respect to the parameter $\lambda_i$ for $i=1,\dots,d$. 
Since an inverse of the matrix $\Lambda_0$ is $\Lambda_0^{-1}=\text{diag}\{1/\lambda_1,\dots,1/\lambda_d\}$, it is clear that
\begin{eqnarray}\label{08012}
\frac{\partial (\pi-\pi_0)'(\Lambda_0^{-1}\otimes \Sigma^{-1})(\pi-\pi_0)}{\partial \lambda_i}&=&-\frac{1}{\lambda_i^2}\mathrm{tr}\big\{(\pi-\pi_0)(\pi-\pi_0)'(\mathsf{E}_{ii}^d\otimes \Sigma^{-1})
\big\},
\end{eqnarray}
where $\mathsf{E}_{ij}^d$ is a $(d\times d)$ matrix and its $(i,j)$--th element equals 1 and others 0 and for generic square matrix $A$, $\text{tr}(A)$ denotes trace of the matrix $A$. In general case, to obtain parameter estimation of the matrix $\Lambda_0$, one may use the following partial derivative 
\begin{equation}\label{08013}
\frac{\partial \mathrm{vec}\big(\Lambda_0^{-1}\otimes\Sigma^{-1}\big)}{\partial \mathrm{vec}(\Lambda_0)'}=-\Big[\mathrm{vec}\big(\Lambda_0^{-1}\mathsf{E}_{11}^d\Lambda_0^{-1}\otimes\Sigma^{-1}\big):\dots:\mathrm{vec}\big(\Lambda_0^{-1}\mathsf{E}_{dd}^d\Lambda_0^{-1}\otimes\Sigma^{-1}\big)\Big].
\end{equation}
It follows from mean and covariance matrix of the random vector $\pi$, given in equation \eqref{08009} that
\begin{equation}\label{08014}
\mathbb{E}\Big[\Big(\pi-\pi_0^{[k+1]}\Big)\Big(\pi-\pi_0^{[k+1]}\Big)'\Big|\Sigma,\mathcal{F}_T;\theta^{[k]}\Big]=\Big(\Lambda_{0|T}^{[k]}\otimes \Sigma\Big)+\Theta_{|T}^{[k]},
\end{equation}
where 
\begin{equation}\label{08015}
\Theta_{|T}^{[k]}:=\Big(\pi_{0|T}^{[k]}-\pi_0^{[k+1]}\Big)\Big(\pi_{0|T}^{[k]}-\pi_0^{[k+1]}\Big)'
\end{equation}
is a positive semi--definite $(d\times d)$ matrix,
\begin{equation}\label{08016}
\Lambda_{0|T}^{[k]}:=\Big(\mathsf{Y}^\circ(\mathsf{Y}^\circ)'+\big(\Lambda_0^{[k]}\big)^{-1}\Big)^{-1}
\end{equation}
is a positive semi--definite $(d\times d)$ matrix and
\begin{equation}\label{08017}
\pi_{0|T}^{[k]}:=\Big(\Lambda_{0|T}^{[k]}\mathsf{Y}^\circ\otimes I_n\Big)y+\Big(\Lambda_{0|T}^{[k]}\big(\Lambda_0^{[k]}\big)^{-1}\otimes I_n\Big)\pi_0^{[k]}
\end{equation}
is an $([nd]\times 1)$ vector. $\Lambda_{0|T}^{[k]}$ and $\pi_{0|T}^{[k]}$ are Bayesian estimators at iteration $k$ of the parameter matrix $\Lambda_0$ and the parameter vector $\pi$ at iteration $k$, respectively. Note that the vector $\pi_{0|T}^{[k]}$ does not depend on the random covariance matrix $\Sigma$. According to the iterated expectation formula and expectation formula of Wishart distributed random matrix $\Sigma^{-1}$, we have that
\begin{eqnarray}\label{08018}
\Psi_{i|T}^{[k]} &:=&\mathbb{E}\Big[\Big(\pi-\pi_0^{[k+1]}\Big)\Big(\pi-\pi_0^{[k+1]}\Big)'(\mathsf{E}_{ii}^d\otimes \Sigma^{-1})\Big|\mathcal{F}_T;\theta^{[k]}\Big]\nonumber\\
&=&\Big(\Lambda_{0|T}^{[k]}\mathsf{E}_{ii}^d\otimes I_n\Big)+\Big(\nu_0^{[k]}+T\Big)\Theta_{|T}^{[k]}\Big(\mathsf{E}_{ii}^d\otimes \big(V_0^{[k]}+B_T^{[k]}\big)^{-1}\Big),
\end{eqnarray}
where the matrix $B_T^{[k]}$ equals
\begin{eqnarray}\label{ad.002}
B_T^{[k]}:=\Big(y^{\circ}-(\pi_0^{\circ})^{[k]}\mathsf{Y}^\circ\Big)\Big(I_T+(\mathsf{Y}^\circ)'\Lambda_0^{[k]}\mathsf{Y}^\circ\Big)^{-1}\Big(y^{\circ}-(\pi_0^{\circ})^{[k]}\mathsf{Y}^\circ\Big)'.
\end{eqnarray}
It can be shown that the matrix $\Psi_{i|T}^{[k]}$ is a positive semi--definite matrix. Consequently, as $\ln|\Lambda_0|=\sum_{i=1}^d\ln(\lambda_i)$, for $i=1,\dots,d$, an estimator at iteration $(k+1)$ of the parameter $\lambda_i$ is given by
\begin{eqnarray}\label{08019}
\lambda_i^{[k+1]}=\frac{1}{n}\text{tr}\Big\{\Psi_{i|T}^{[k]}\Big\}.
\end{eqnarray}
Because $\Psi_{i|T}^{[k]}$ is a positive semi--definite matrix, $\lambda_i^{[k+1]}$ takes a non--negative value. For $i=1,\dots,d$, we collect $\lambda_i^{[k+1]}$ into a matrix $\Lambda_0^{[k+1]}:=\text{diag}\Big\{\lambda_1^{[k+1]},\dots,\lambda_d^{[k+1]}\Big\}$.

Second, we consider partial derivative from the objective function $\mathcal{L}$ with respect to the parameter vector $\pi_0$. The partial derivative is given by
\begin{equation}\label{08020}
\frac{\partial \mathcal{L}}{\partial \pi_0}=\mathbb{E}\big[(\Lambda_0^{-1}\otimes \Sigma^{-1})(\pi-\pi_0)\big|\mathcal{F}_T;\theta^{[k]}\big].
\end{equation}
According to the iterated expectation formula and equations \eqref{08009} and \eqref{08010}, one obtains estimator at iteration $(k+1)$ of the parameter vector $\pi_0$
\begin{equation}\label{08021}
\pi_0^{[k+1]}=\pi_{0|T}^{[k]}.
\end{equation}

Third, we consider partial derivative from the objective function $\mathcal{L}$ with respect to the parameter matrix $V_0$. The partial derivative is given by
\begin{equation}\label{08022}
\frac{\partial \mathcal{L}}{\partial V_0}=\mathbb{E}\bigg[\frac{\nu_0}{2}V_0^{-1}-\frac{1}{2}\Sigma^{-1}\bigg|\mathcal{F}_T;\theta^{[k]}\bigg].
\end{equation}
Consequently, we have that
\begin{equation}\label{08023}
V_0^{[k+1]}=\frac{\nu_0^{[k+1]}}{\nu_0^{[k]}+T}\Big(V_0^{[k]}+B_T^{[k]}\Big).
\end{equation}

Fourth, we consider partial derivative from the objective function $\mathcal{L}$ with respect to the parameter $\nu_0$. The partial derivative is given by
\begin{eqnarray}\label{08024}
\frac{\partial \mathcal{L}}{\partial \nu_0}=\mathbb{E}\bigg[\frac{1}{2}\ln|\Sigma^{-1}|+\frac{1}{2}\ln|V_0|-\frac{1}{2}\psi_n\bigg(\frac{\nu_0}{2}\bigg)-\frac{n}{2}\ln(2)\bigg|\mathcal{F}_T;\theta^{[k]}\bigg],
\end{eqnarray}
where $\psi_n(\cdot)$ is the multivariate digamma function, which is defined by derivative of log of the multivariate gamma function. As a result, since $\mathbb{E}\big[\ln|\Sigma^{-1}|\big|\mathcal{F}_T\big]=\psi_n(\frac{\nu_0+T}{2})+n\ln(2)-\ln|V_0+B_T|$, see \citeA{Nguyen23}, one gets that
\begin{equation}\label{08025}
\psi_n\bigg(\frac{\nu_0^{[k]}+T}{2}\bigg)-\psi_n\bigg(\frac{\nu_0^{[k+1]}}{2}\bigg)-\ln\Big|V_0^{[k]}+B_T^{[k]}\Big|+\ln\Big|V_0^{[k+1]}\Big|=0.
\end{equation}
If we substitute equation \eqref{08023} into the above equation, then we have that
\begin{equation}\label{ad.001}
\psi_n\bigg(\frac{\nu_0^{[k]}+T}{2}\bigg)-\psi_n\bigg(\frac{\nu_0}{2}\bigg)-\ln\Big(\nu_0^{[k]}+T\Big)+\ln(\nu_0)=0.
\end{equation}
As a result, to obtain an estimator at iteration $(k+1)$ of the parameter $\nu_0$, one has to solve the above nonlinear equation for $\nu_0$.
 
In the following algorithm, we give EM algorithm for parameters of conditional matrix variate $t$ density function

\begin{algorithm}\label{algo01}
\normalfont \textbf{(EM Estimation of General Type I Conditional Matrix Variate $t$ Distribution).}
\begin{itemize}
\item[(1)] Set $k=0$ and initial value of the parameter vector $\theta^{[0]}=\big(\pi_0^{[0]},\Lambda_0^{[0]},\nu_0^{[0]},V_0^{[0]}\big)$. 
\item[(2)] Calculate $\pi_0^{[k+1]}$ using equation \eqref{08021}.
\item[(3)] For $i=1,\dots,d$, calculate $\lambda_i^{[k+1]}$ using equation \eqref{08019} and collect $\lambda_i^{[k+1]}$ into a diagonal matrix $\Lambda_0^{[k+1]}:=\text{diag}\Big\{\lambda_1^{[k+1]},\dots,\lambda_d^{[k+1]}\Big\}.$
\item[(4)] To obtain $\nu_0^{[k+1]}$, solve nonlinear equation \eqref{ad.001} for $\nu_0$.
\item[(5)] Calculate $V_0^{[k+1]}$ using equation \eqref{08023}.
\item[(6)] Increase iteration count $k=k+1$ and go to step (2).
\end{itemize}
\end{algorithm}

\subsection{Type I Conditional Matrix Variate $t$ Distribution with Minnesota Prior}

In this subsection, we follow \citeA{Battulga24g} and we develop EM algorithm for parameters of type I conditional matrix variate $t$ distribution with Minnesota prior. In practice, one usually adopts the Minnesota prior to estimating the parameters of the VAR$(p)$ process. The first version of  Minnesota prior was introduced by \citeA{Litterman79}. Also, \citeA{Banbura10} used Minnesota prior for large Bayesian VAR and showed that the forecast of large Bayesian VAR is better than small Bayesian VAR. However, there are many different variants of the Minnesota prior, we consider a prior, which is included in \citeA{Miranda18}. The idea of Minnesota prior is that it shrinks diagonal elements of the matrix $A_1$ toward $\phi_i$ and off--diagonal elements of $A_1$ and all elements of other matrices $A_0,A_2,\dots,A_p$ toward 0, where $\phi_i$ is 0 for a stationary variable $y_{i,t}$ and 1 for a variable with unit root $y_{i,t}$. However, we adopt a different prior condition for the random coefficient matrix $A_0$.  Without loss of generality let us assume that there are $m$ $(m=0,\dots,n)$ stationary variables and the stationary variables are placed on the first $m$ components of the process $y_t$. For the prior, it is assumed that conditional on $\Sigma$ and $\mathcal{F}_0$, $A_0,A_1,\dots,A_p$ are jointly normally distributed, and for $(i,j)$--th element of the matrix $A_{\ell}$ $(\ell=0,\dots,p)$, it holds that for $i=1,\dots,n$ and $j=1,\dots,l$,
\begin{equation}\label{08026}
\mathbb{E}\big((A_0)_{i,j}\big|\Sigma,\mathcal{F}_0\big)=
\begin{cases}
C_{i,j} & \text{if}~~~i=1,\dots,m\\
0 & \text{if}~~~i=m+1,\dots,n
\end{cases}
\end{equation}
and
\begin{equation}\label{08027}
\text{Var}\big((A_0)_{i,j}\big|\Sigma,\mathcal{F}_0\big)=(\sigma_i/\varepsilon_j)^2
\end{equation}
and for $i,j=1,\dots,n$,
\begin{equation}\label{08028}
\mathbb{E}\big((A_{\ell})_{i,j}\big|\Sigma,\mathcal{F}_0\big)=\begin{cases}
\phi_i & \text{if}~~~i=j,~\ell=1,\\
0 & \text{if}~~~\text{otherwise}
\end{cases},~~~\text{for}~\ell=1,\dots,p
\end{equation}
and
\begin{equation}\label{08029}
\text{Var}\big((A_{\ell})_{i,j}\big|\Sigma,\mathcal{F}_0\big)=\begin{cases}
\displaystyle \bigg(\frac{\sigma_i}{\alpha\ell^{\beta}\gamma_i}\bigg)^2 & \text{if}~~~i=j,\\
\displaystyle \bigg(\frac{\sigma_i}{\alpha\ell^{\beta}\gamma_j}\bigg)^2 & \text{if}~~~\text{otherwise}
\end{cases}~~~\text{for}~\ell=1,\dots,p,
\end{equation}
where $C_{i,j}$ is $(i,j)$--th element of an $(n\times m)$ parameter matrix $C_m$ and $\sigma_i^2$ is an $(i,i)$--th element of the random covariance matrix $\Sigma$. It follows from equation \eqref{08026} that the expectation of $(A_0)_{i,j}$ equals $C_{i,j}$ for stationary variable $y_{i,t}$ and zero for non stationary variable with unit root $y_{i,t}$. Let us introduce an $(n\times l)$ matrix $\bar{C}_m:=[C_m:0_{[n\times (l-m)]}]$. A small $\varepsilon_i^2$ corresponds to an uninformative diffuse prior for $(A_0)_{i,j}$, the parameter $\alpha$ controls the overall tightness of the prior distribution, the parameter $\beta$ controls amount of information prior information at higher lags, and $\gamma_i$ is a scaling parameter, see \citeA{Miranda18}. Thus, the factor $1/\ell^{2\beta}$ represents a rate at which prior variance decreases with increasing lag length. 

According to \citeA{Banbura10}, it can be shown that the following equation satisfies the prior conditions \eqref{08026}--\eqref{08029}
\begin{equation}\label{08030}
\hat{y}^\circ=\Pi\hat{\mathsf{Y}}^\circ+\hat{\xi}^\circ,
\end{equation}
where $\hat{y}^\circ$ and $\hat{\mathsf{Y}}^\circ$ are $(n\times d)$ and $(d\times d)$ matrices of dummy variables and are defined by
\begin{equation}\label{08031}
\hat{y}^\circ:=\big[\text{diag}\{1-\phi_1,\dots,1-\phi_n\}\bar{C}_m\text{diag}\{\varepsilon_1,\dots,\varepsilon_l\}:\alpha\text{diag}\{\phi_1\gamma_1,\dots,\phi_n\gamma_n\}:0_{[n\times n(p-1)]}\big]
\end{equation}
and
\begin{equation}\label{08032}
\hat{\mathsf{Y}}^\circ:=\begin{bmatrix}
\text{diag}\{\varepsilon_1,\dots,\varepsilon_l\} & 0_{[l\times np]}\\
0_{[np\times l]} & \alpha\big(J_\beta\otimes\text{diag}\{\gamma_1,\dots,\gamma_n\}\big)
\end{bmatrix}
\end{equation}
with $J_\beta:=\text{diag}\{1^{\beta},\dots,p^{\beta}\}$, respectively, and $\hat{\xi}^\circ:=[\xi_1:\dots:\xi_d]$ is an $(n\times d)$ matrix of residual process. Note that one can add constraints for elements of the coefficient matrix $\Pi$ to the matrices of dummy variables. It is worth mentioning that the matrices of dummy variables $\hat{y}_t$ and $\hat{\mathsf{Y}}_t$ should not depend on the covariance matrix $\Sigma$. If the dummy variables depend on the covariance matrix, an OLS estimator, and matrix $\Lambda_{0}$ depend on the covariance matrix $\Sigma$, see below. Consequently, in this case, one can not use the results of Proposition \ref{prop01}. For this reason, we choose the prior condition \eqref{08026}--\eqref{08029}. Equation \eqref{08030} can be written by
\begin{equation}\label{08033}
\hat{y}=\big((\hat{\mathsf{Y}}^\circ)'\otimes I_n\big)\pi+\hat{\xi},
\end{equation}
where $\hat{y}$ and $\hat{\xi}$ are $([nd]\times 1)$ vectors and are vectorizations of the matrix of dummy variables $\hat{y}^\circ$ and matrix of the residual process $\hat{\xi}^\circ$, respectively, i.e., $\hat{y}:=\text{vec}(\hat{y}^\circ)$ and $\hat{\xi}:=\text{vec}(\hat{\xi}^\circ)$. It follows from equation \eqref{08033} that
\begin{equation}\label{08034}
\pi\overset{d}{=}\big(((\hat{\mathsf{Y}}^\circ (\hat{\mathsf{Y}}^\circ)')^{-1}\hat{\mathsf{Y}}^\circ)\otimes I_n\big)\hat{y}+\big(((\hat{\mathsf{Y}}^\circ (\hat{\mathsf{Y}}^\circ)')^{-1}\hat{\mathsf{Y}}^\circ)\otimes I_n\big)\hat{\xi},
\end{equation}
where $d$ denotes equal distribution. It should be noted that the first term of the right--hand side of the above equation is a vecorization of the ordinary least square (OLS) estimator of the coefficient matrix $\Pi$, namely, 
\begin{eqnarray}\label{08035}
\pi_0^\circ&:=&\hat{y}_t^\circ(\hat{\mathsf{Y}}^\circ)'(\hat{\mathsf{Y}}^\circ(\hat{\mathsf{Y}}^\circ)')^{-1}\nonumber\\
&=&\big[\text{diag}\{1-\phi_1,\dots,1-\phi_n\}\bar{C}_m:\text{diag}\{\phi_1,\dots,\phi_n\}:0_{[n\times n(p-1)]}\big].
\end{eqnarray}
Note that the OLS estimator of the coefficient matrix $\Pi$ are same as the prior conditions \eqref{08026} and \eqref{08028}. Consequently, conditional on $\Sigma$ and $\mathcal{F}_0$, a distribution of the coefficient vector $\pi$ is given by
\begin{equation}\label{08036}
\pi~|~\Sigma,\mathcal{F}_0\sim \mathcal{N}\Big(\pi_0,\big(\Lambda_0\otimes \Sigma\big)\Big).
\end{equation}
where $\Lambda_0:=(\hat{\mathsf{Y}}^\circ(\hat{\mathsf{Y}}^\circ)')^{-1}$ is a $(d\times d)$ diagonal matrix and its inverse equals
\begin{equation}\label{08037}
\Lambda_0^{-1}:=\hat{\mathsf{Y}}^\circ(\hat{\mathsf{Y}}^\circ)'=\begin{bmatrix}
\text{diag}\{\varepsilon_1^2,\dots,\varepsilon_l^2\} & 0_{[l\times np]}\\
0_{[np\times l]} & \text{diag}\{1^{2\beta},\dots,p^{2\beta}\}\otimes\text{diag}\{\alpha^2\gamma_1^2,\dots,\alpha^2\gamma_n^2\}
\end{bmatrix}.
\end{equation}
Since the matrix $\Lambda_0^{-1}$ is a diagonal matrix, its determinant, which appears in the objective function \eqref{08011} is
\begin{equation}\label{08038}
\ln|\Lambda_0^{-1}|=2\sum_{j=1}^l\ln(\varepsilon_j)+2np\ln(\alpha)+2\beta\sum_{\ell=1}^p\ln(\ell)+2p\sum_{i=1}^n\ln(\gamma_i).
\end{equation}
Let for $m=1,\dots,l$, $c_m:=\text{vec}(C_m)$ be a $(nm\times 1)$ vectorization of the parameter matrix $C_m$. Then, partial derivative from the objective function $\mathcal{L}$ with respect to the parameter $c$ is
\begin{eqnarray}\label{08039}
\frac{\partial \mathcal{L}}{\partial c'}&=&\mathbb{E}\big[(\pi-\pi_0)'(\Lambda_0^{-1}\otimes \Sigma^{-1})J_m'\big|\mathcal{F}_T;\theta^{[k]}\big]\nonumber\\
&=&\mathbb{E}\big[(\pi-\pi_0)'J_m'\big(D_m(\varepsilon)\otimes \Sigma^{-1}\big)\big|\mathcal{F}_T;\theta^{[k]}\big],
\end{eqnarray}
where $J_m:=\big[I_{mn}:0_{[mn\times n^2p]}\big]$ is an $(mn\times dn)$ matrix and $D_m(\varepsilon):=\text{diag}\{\varepsilon_1^2,\dots,\varepsilon_m^2\}$ is an $(m\times m)$ diagonal matrix. Then, since $J_m(\pi-\pi_0)=J_m\pi-c_m$, we get an estimator at iteration $(k+1)$ of the parameter vector $c_m$
\begin{equation}\label{08040}
c_m^{[k+1]}:=J_m\pi_{0|T}^{[k]},
\end{equation}
where $\pi_{0|T}^{[k]}$ is given by equation \eqref{08017}. Note that if all variables of the process $y_t$ are unit root processes ($m=0$), then one does not need a parameter estimation of the matrix $C_m$. To obtain estimators at iteration $(k+1)$ of the parameters $\alpha$, $\beta$, $\gamma_i$, and $\varepsilon_j$ for $i=1,\dots,n$ and $j=1,\dots,l$, we define the following $(d\times d)$ matrices:
\begin{equation}\label{08041}
\Delta_{\varepsilon_j}:=\begin{bmatrix}
 \mathsf{E}_{jj}^l & 0_{[l\times np]}\\
0_{[np\times l]} & 0_{[np\times np]}
\end{bmatrix},
\end{equation}
\begin{equation}\label{08042}
\Delta_\alpha^{[k]}:=\begin{bmatrix}
0_{l\times l} & 0_{[l\times np]}\\
0_{[np\times l]} & \text{diag}\big\{1^{2\beta^{[k]}},\dots,p^{2\beta^{[k]}}\big\}\otimes\text{diag}\big\{\big(\gamma_1^{[k]}\big)^2,\dots,\big(\gamma_n^{[k]}\big)^2\big\}
\end{bmatrix},
\end{equation}
\begin{equation}\label{08043}
\Delta_{\beta}^{[k]}:=\big(\alpha^{[k]}\big)^2\begin{bmatrix}
0_{l\times l} & 0_{[l\times np]}\\
0_{[np\times l]} & \text{diag}\big\{1^{2\beta}\ln(1),\dots,p^{2\beta}\ln(p)\big\}\otimes\text{diag}\big\{\big(\gamma_1^{[k]}\big)^2,\dots,\big(\gamma_n^{[k]}\big)^2\big\}
\end{bmatrix},
\end{equation}
and
\begin{equation}\label{08044}
\Delta_{\gamma_i}^{[k]}:=\big(\alpha^{[k]}\big)^2\begin{bmatrix}
0_{l\times l} & 0_{[l\times np]}\\
0_{[np\times l]} & \text{diag}\big\{1^{2\beta^{[k]}}\ln(1),\dots,p^{2\beta^{[k]}}\ln(p)\big\}\otimes\mathsf{E}_{ii}^n
\end{bmatrix}.
\end{equation}
It follows from determinant equation \eqref{08038} and the objective function \eqref{08011} that similarly to equation \eqref{08019}, one obtains that for $i=1,\dots,n$ and $j=1,\dots,l$, estimators at iteration $(k+1)$ of the parameters $\varepsilon_j$, $\alpha$, and $\gamma_i$ are given by
\begin{eqnarray}\label{08045}
\Big(\varepsilon_j^{[k+1]}\Big)^2:=\frac{n}{\text{tr}\Big\{\Big(\Lambda_{0|T}^{[k]}\Delta_{\varepsilon_j}\otimes I_n\Big)+\Big(\nu_0^{[k]}+T\Big)\Theta_{|T}^{[k]}\Big(\Delta_{\varepsilon_j}\otimes \big(V_0^{[k]}+B_T^{[k]}\big)^{-1}\Big)\Big\}},
\end{eqnarray}
\begin{eqnarray}\label{08046}
\Big(\alpha^{[k+1]}\Big)^2:=\frac{n^2p}{\text{tr}\Big\{\Big(\Lambda_{0|T}^{[k]}\Delta_\alpha^{[k]}\otimes I_n\Big)+\Big(\nu_0^{[k]}+T\Big)\Theta_{|T}^{[k]}\Big(\Delta_\alpha^{[k]}\otimes \big(V_0^{[k]}+B_T^{[k]}\big)^{-1}\Big)\Big\}},
\end{eqnarray}
and
\begin{eqnarray}\label{08047}
\Big(\gamma_i^{[k+1]}\Big)^2:=\frac{np}{\text{tr}\Big\{\Big(\Lambda_{0|T}^{[k]}\Delta_{\gamma_i}^{[k]}\otimes I_n\Big)+\Big(\nu_0^{[k]}+T\Big)\Theta_{|T}^{[k]}\Big(\Delta_{\gamma_i}^{[k]}\otimes \big(V_0^{[k]}+B_T^{[k]}\big)^{-1}\Big)\Big\}},
\end{eqnarray}
where $\Theta_{|T}^{[k]}$ is calculated via equations \eqref{08015}, \eqref{08035}, and \eqref{08040}.
For $p\geq 2$, an estimator at iteration $(k+1)$ of the parameter $\beta$, $\beta^{[k+1]}$ is obtained from the following nonlinear equation
\begin{eqnarray}\label{08048}
\sum_{\ell=1}^p\ln(\ell)=\frac{1}{n}\text{tr}\Big\{\Big(\Lambda_{0|T}^{[k]}\Delta_{\beta}^{[k]}\otimes I_n\Big)+\Big(\nu_0^{[k]}+T\Big)\Theta_{|T}^{[k]}\Big(\Delta_{\beta}^{[k]}\otimes \big(V_0^{[k]}+B_T^{[k]}\big)^{-1}\Big)\Big\}.
\end{eqnarray}
It should be noted that it is not difficult to show that the right--hand sides of equations \eqref{08045}--\eqref{08048} take non negative values. Consequently, an EM algorithm for parameters of conditional matrix variate $t$ distribution with Minnesota prior is given by the following algorithm.

\begin{algorithm}\label{algo02}
\normalfont \textbf{(EM Estimation of Type I Conditional Matrix Variate $t$ Distribution with Minnesota Prior).}
\begin{itemize}
\item[(1)] Set $k=0$ and initial value of the parameter vector 
\begin{equation}\label{08049}
\theta^{[0]}=\Big(C_m^{[0]},\varepsilon_1^{[0]},\dots,\varepsilon_l^{[0]},\alpha^{[0]},\beta^{[0]},\gamma_1^{[0]},\dots,\gamma_n^{[0]},\nu_0^{[0]},V_0^{[0]}\Big).
\end{equation}
\item[(2)] If $1\leq m\leq n$, calculate $C_m^{[k+1]}$ using equation \eqref{08040} and set $\bar{C}_m^{[k+1]}=\big[C_m^{[k+1]}:0\big]$. Otherwise set $\bar{C}_m^{[k+1]}=0$. 
\item[(3)] For $i=1,\dots,n$ and $j=1,\dots,l$, calculate $\Big(\varepsilon_j^{[k+1]}\Big)^2$, $\Big(\alpha^{[k+1]}\Big)^2$, and $\Big(\gamma_i^{[k+1]}\Big)^2$ using equations \eqref{08045}--\eqref{08047}.
\item[(4)] For $p\geq 2$, to obtain $\beta^{[k+1]}$, solve nonlinear equation \eqref{08048} for $\beta$.
\item[(5)] To obtain $\nu_0^{[k+1]}$, solve nonlinear equation \eqref{08025} for $\nu_0$.
\item[(6)] Calculate $V_0^{[k+1]}$ using equation \eqref{08023}.
\item[(7)] Increase iteration count $k=k+1$ and go to step (2).
\end{itemize}
\end{algorithm}

\section{EM Estimation of Type II Conditional Matrix Variate $t$ Distribution}

Let us reconsider a BVAR($p$) process, given in equation \eqref{08001},  namely,  
\begin{equation}\label{08050}
y_t=\Pi_t\mathsf{Y}_{t}+\xi_t=(\mathsf{Y}_{t}'\otimes I_n)\pi_t+\xi_t, ~~~t=1,\dots,T,
\end{equation}
where $\pi_t:=\text{vec}(\Pi_t)$ is an $(nd\times 1)$ vectorization of the random coefficient matrix $\Pi_t$.  For the random coefficient vectors and covariance matrices, we assume the following assumption holds.

\begin{assumption}\label{ass01}
Conditional on the initial information $\mathcal{F}_0$, the random coefficient vectors and covariance matrices $(\pi_1,\Sigma_1),\dots,(\pi_T,\Sigma_T)$ are independent and identically distributed.
\end{assumption}

Note that from the assumption, we can conclude that conditional on initial information $\mathcal{F}_0$, for each $t=2,\dots,T$, $(\pi_t,\Sigma_t)$ is independent of a random vector $(y_1',\dots,y_{t-1}')'$. Now we define distributions of the random coefficient vector $\pi_t$ and covariance matrix $\Sigma_t$. We suppose that conditional on the initial information $\mathcal{F}_0$, a distribution of the random covariance matrix $\Sigma_t$ is given by 
\begin{equation}\label{08051}
\Sigma_t~|~\mathcal{F}_0\sim \mathcal{IW}(\nu_0,V_0).
\end{equation}
where $\nu_0>n-1$ is a degrees of freedom and $V_0$ is a positive definite scale matrix. Hence, a distribution of the residual vector $\xi_t$ equals 
\begin{equation}\label{08052}
\xi_t~|~\Sigma_t,\mathcal{F}_0\sim \mathcal{N}\big(0,\Sigma_t\big).
\end{equation}
Also, we suppose that conditional on the covariance matrix $\Sigma_t$ and initial information $\mathcal{F}_0$, a distribution of the random coefficient vector $\pi_t$ is given by 
\begin{equation}\label{08053}
\pi_t~|~\Sigma_t,\mathcal{F}_0\sim \mathcal{N}\Big(\pi_0,\Lambda_0\otimes \Sigma_t\Big),
\end{equation}
where $\pi_0$ is an $(nd\times 1)$ vector and $\Lambda_0$ is a symmetric positive definite $(d\times d)$ matrix. Then, the following Proposition holds.

\begin{proposition}\label{prop02}
Let for $t=1,\dots,T$, $\pi_t~|~\Sigma_t,\mathcal{F}_0\sim \mathcal{N}\big(\pi_0,\Lambda_0\otimes \Sigma_t\big)$, $\Sigma_t~|~\mathcal{F}_0\sim \mathcal{IW}(\nu_0,V_0)$, and Assumption \ref{ass01} holds. Then, first, conditional on the information $\mathcal{F}_{t-1}$, a joint density function of the random vector $y_t$ is given by
\begin{eqnarray}\label{08054}
f(y_t|\mathcal{F}_{t-1})=\frac{1}{\pi^{n/2}}\frac{\Gamma_{n}\big((\nu_{0}+1)/2\big)|V_{0}|^{\nu_{0}/2}}{\big(1+\mathsf{Y}_t'\Lambda_{0}\mathsf{Y}_t\big)^{n/2}\Gamma_{n}(\nu_{0}/2)\big|\tilde{B}_{t}+V_{0}\big|^{(\nu_{0}+1)/2}},
\end{eqnarray}
where
\begin{eqnarray}\label{08055}
\tilde{B}_{t}:=\frac{1}{1+\mathsf{Y}_t'\Lambda_{0}\mathsf{Y}_t}\big(y_t-\pi_{0}^{\circ}\mathsf{Y}_t\big)\big(y_t-\pi_{0}^{\circ}\mathsf{Y}_t\big)'
\end{eqnarray}
is a positive semi--define $(n\times n)$ matrix. Second, conditional on the random covariance matrix $\Sigma_t$ and information $\mathcal{F}_t$, a joint density function of the random coefficient vector $\pi_t$ is given by
\begin{eqnarray}\label{08056}
f(\pi_t|\Sigma_t,\mathcal{F}_t)= \frac{1}{(2\pi)^{nd/2}|\tilde{\Lambda}_{0|t}|^{n/2} |\Sigma|^{d/2}} \exp\bigg\{-\frac{1}{2}\Big(\pi_t-\tilde{\pi}_{0|t}\Big)'\big(\tilde{\Lambda}_{0|t}^{-1}\otimes \Sigma_t^{-1}\big)\Big(\pi_t-\tilde{\pi}_{0|t}\Big)\bigg\},
\end{eqnarray}
where $\tilde{\Lambda}_{0|t}^{-1}:=\mathsf{Y}_t\mathsf{Y}_t+\Lambda_0^{-1}$ is a $(d\times d)$ matrix and $\tilde{\pi}_{0|t}:=\big((\tilde{\Lambda}_{0|T}\mathsf{Y})\otimes I_n\big)y_t+\big((\tilde{\Lambda}_{0|t}\Lambda_0^{-1})\otimes I_n\big)\pi_0$ is an $([nd]\times 1)$ vector. Finally, conditional on the information $\mathcal{F}_t$, a joint density function of the random coefficient matrix $\Sigma$ is given by
\begin{eqnarray}\label{08057}
f(\Sigma_t|\mathcal{F}_t)&=&\frac{\big|\tilde{B}_t+V_0\big|^{(\nu_0+1)/2}}{\Gamma_n\big((\nu_0+1)/2\big)2^{n(\nu_0+1)/2}}|\Sigma_t|^{-(\nu_0+n+2)/2}\exp\bigg\{-\frac{1}{2}\mathrm{tr}\Big(\big(\tilde{B}_t+V_0\big)\Sigma_t^{-1}\Big)\bigg\}.
\end{eqnarray}
\end{proposition}

We refer to distribution function, which is given in equation \eqref{08054} as type II conditional matrix variate $t$ distribution function. In this section, we introduce an EM algorithm to estimate the parameters of the distribution function. Similarly to equation \eqref{08011}, an conditional expectation (objective function) in E--Step is given by
\begin{eqnarray}\label{08058}
\mathcal{L}&=&\mathbb{E}\bigg[-\frac{nT(1+d)}{2}\ln(2\pi)+\sum_{t=1}^T\frac{d+\nu_{0}+n+2}{2}\ln|\Sigma_t^{-1}|\nonumber\\
&-&\frac{1}{2}\sum_{t=1}^T\big(y_t-(\mathsf{Y}_t'\otimes I_n)\pi_t\big)'\Sigma_t^{-1}\big(y_t-(\mathsf{Y}_t'\otimes I_n)\pi_t\big)\nonumber\\
&-&\frac{nT}{2}\ln|\Lambda_{0}|-\frac{1}{2}\sum_{t=1}^T(\pi_t-\pi_{0})'(\Lambda_{0}^{-1}\otimes \Sigma_t^{-1})(\pi_t-\pi_{0})\\
&+&\frac{\nu_{0}T}{2}\ln|V_{0}|-\sum_{t=1}^T\ln\bigg(\Gamma_n\bigg(\frac{\nu_{0}}{2}\bigg)\bigg)-\sum_{t=1}^T\frac{n\nu_{0}}{2}\ln(2)-\frac{1}{2}\sum_{t=1}^T\mathrm{tr}\big(V_{0}\Sigma_t^{-1}\big)\bigg|\mathcal{F}_T;\theta^{[k]}\bigg]\nonumber.
\end{eqnarray}
In M--Step, to obtain ML estimators at iteration $(k+1)$ of the parameters of the type II conditional matrix variate $t$ distribution, we need the following Lemma.
\begin{lemma}\label{lem01}
Under Assumption \ref{ass01}, conditional on the full information $\mathcal{F}_T$, a joint density function of coefficient vector $\pi_t$ and covariance matrix $\Sigma_t$ is given by
\begin{equation}\label{08059}
f(\pi_t,\Sigma_t|\mathcal{F}_T)=f(\pi_t,\Sigma_t|\mathcal{F}_t).
\end{equation}
\end{lemma}

By using Proposition \ref{prop02}, Lemma \ref{lem01}, and ideas in subsection 2.1, one can arrive the following EM Algorithm, which estimates parameters of the general type II conditional matrix variate $t$ distribution.
\begin{algorithm}\label{algo03}
\normalfont \textbf{(EM Estimation of General Type II Conditional Matrix Variate $t$ Distribution).}
\begin{itemize}
\item[(1)] Set $k=0$ and initial value of the parameter vector $\theta^{[0]}=\big(\pi_0^{[0]},\Lambda_0^{[0]},\nu_0^{[0]},V_0^{[0]}\big)$. 
\item[(2)] Calculate $\pi_0^{[k+1]}$ using equation 
\begin{eqnarray}\label{08060}
\pi_0^{[k+1]}&=&\bigg(\sum_{t=1}^T\Big(I_d\otimes \big(V_0^{[k]}+\tilde{B}_t^{[k]}\big)^{-1}\Big)\bigg)^{-1}\sum_{t=1}^T\Big(I_d\otimes \big(V_0^{[k]}+\tilde{B}_t^{[k]}\big)^{-1}\Big)\tilde{\pi}_{0|t}^{[k]},
\end{eqnarray}
where the matrix and vector are given by
\begin{eqnarray}\label{08055}
\tilde{B}_{t}^{[k]}:=\frac{1}{1+\mathsf{Y}_t'\Lambda_{0}^{[k]}\mathsf{Y}_t}\Big(y_t-\big(\pi_{0}^{\circ}\big)^{[k]}\mathsf{Y}_t\Big)\Big(y_t-\big(\pi_{0}^{\circ}\big)^{[k]}\mathsf{Y}_t\Big)'
\end{eqnarray}
and
\begin{equation}\label{08063}
\tilde{\pi}_{0|t}^{[k]}:=\Big(\tilde{\Lambda}_{0|t}^{[k]}\mathsf{Y}_t\otimes I_n\Big)y_t+\Big(\tilde{\Lambda}_{0|t}^{[k]}\big(\Lambda_0^{[k]}\big)^{-1}\otimes I_n\Big)\pi_0^{[k]}.
\end{equation}
\item[(3)] For $i=1,\dots,d$, calculate $\lambda_i^{[k+1]}$ using equation 
\begin{eqnarray}\label{08061}
\lambda_i^{[k+1]}=\frac{1}{nT}\sum_{t=1}^T\text{tr}\Big\{\Big(\tilde{\Lambda}_{0|t}^{[k]}\mathsf{E}_{ii}^d\otimes I_n\Big)+\Big(\nu_0^{[k]}+1\Big)\tilde{\Theta}_{|t}^{[k]}\Big(\mathsf{E}_{ii}^d\otimes \big(V_0^{[k]}+\tilde{B}_t^{[k]}\big)^{-1}\Big)\Big\},
\end{eqnarray}
where the matrices are given by
\begin{equation}\label{08062}
\tilde{\Theta}_{|t}^{[k]}:=\Big(\tilde{\pi}_{0|t}^{[k]}-\pi_0^{[k+1]}\Big)\Big(\tilde{\pi}_{0|t}^{[k]}-\pi_0^{[k+1]}\Big)' ~~~\text{and}~~~\tilde{\Lambda}_{0|t}^{[k]}:=\Big(\mathsf{Y}_t\mathsf{Y}_t'+\big(\Lambda_0^{[k]}\big)^{-1}\Big)^{-1}.
\end{equation}
Collect $\lambda_i^{[k+1]}$ into a diagonal matrix $\tilde{\Lambda}_0^{[k+1]}:=\text{diag}\Big\{\lambda_1^{[k+1]},\dots,\lambda_d^{[k+1]}\Big\}.$
\item[(4)] To obtain $\nu_0^{[k+1]}$, solve the following nonlinear equation for $\nu_0$
\begin{eqnarray}\label{08064}
&&\sum_{t=1}^T\Bigg(\psi_n\bigg(\frac{\nu_0^{[k]}+1}{2}\bigg)-\psi_n\bigg(\frac{\nu_0}{2}\bigg)-\ln\Big|V_0^{[k]}+\tilde{B}_t^{[k]}\Big|+\ln(T)\nonumber\\
&&+\ln(\nu_0)-\ln\Big(\nu_0^{[k]}+1\Big)-\ln\bigg|\sum_{t=1}^T\Big(V_0^{[k]}+\tilde{B}_t^{[k]}\Big)^{-1}\bigg|\Bigg)=0.
\end{eqnarray}
\item[(5)] Calculate $V_0^{[k+1]}$ using equation 
\begin{equation}\label{08065}
V_0^{[k+1]}=\frac{T\nu_0^{[k+1]}}{\nu_0^{[k]}+1}\Bigg(\sum_{t=1}^T\Big(V_0^{[k]}+\tilde{B}_t^{[k]}\Big)^{-1}\Bigg)^{-1}.
\end{equation}
\item[(6)] Increase iteration count $k=k+1$ and go to step (2).
\end{itemize}
\end{algorithm}

Now we consider type II conditional matrix variate $t$ distribution  with Minnesota prior. For random coefficient matrices $A_{0,t},A_{1,t},\dots,A_{p,t}$, we assume that same prior conditions hold as subsection 2.2, namely, for $i=1,\dots,n$ and $j=1,\dots,l$,
\begin{equation}\label{08066}
\mathbb{E}\big((A_{0,t})_{i,j}\big|\Sigma_t,\mathcal{F}_0\big)=
\begin{cases}
C_{i,j} & \text{if}~~~i=1,\dots,m\\
0 & \text{if}~~~i=m+1,\dots,n
\end{cases}
\end{equation}
and
\begin{equation}\label{08067}
\text{Var}\big((A_{0,t})_{i,j}\big|\Sigma_t,\mathcal{F}_0\big)=(\sigma_{i,t}/\varepsilon_j)^2
\end{equation}
and for $i,j=1,\dots,n$,
\begin{equation}\label{08068}
\mathbb{E}\big((A_{\ell,t})_{i,j}\big|\Sigma_t,\mathcal{F}_0\big)=\begin{cases}
\phi_i & \text{if}~~~i=j,~\ell=1,\\
0 & \text{if}~~~\text{otherwise}
\end{cases},~~~\text{for}~\ell=1,\dots,p
\end{equation}
and
\begin{equation}\label{08069}
\text{Var}\big((A_{\ell,t})_{i,j}\big|\Sigma_t,\mathcal{F}_0\big)=\begin{cases}
\displaystyle \bigg(\frac{\sigma_{i,t}}{\alpha\ell^{\beta}\gamma_i}\bigg)^2 & \text{if}~~~i=j,\\
\displaystyle \bigg(\frac{\sigma_{i,t}}{\alpha\ell^{\beta}\gamma_j}\bigg)^2 & \text{if}~~~\text{otherwise}
\end{cases}~~~\text{for}~\ell=1,\dots,p,
\end{equation}
where $\sigma_{i,t}^2$ is an $(i,i)$--th element of the random covariance matrix $\Sigma_t$. Then, similarly to the Algorithm \ref{algo02}, one obtains the following EM algorithm, which estimate parameters of type II conditional matrix variate $t$ distribution with Minnesota prior.

\begin{algorithm}\label{algo04}
\normalfont \textbf{(EM Estimation of Type II Conditional Matrix Variate $t$ Distribution with Minnesota Prior).}
\begin{itemize}
\item[(1)] Set $k=0$ and initial value of the parameter vector 
\begin{equation}\label{08070}
\theta^{[0]}=\Big(C_m^{[0]},\varepsilon_1^{[0]},\dots,\varepsilon_l^{[0]},\alpha^{[0]},\beta^{[0]},\gamma_1^{[0]},\dots,\gamma_n^{[0]},\nu_0^{[0]},V_0^{[0]}\Big).
\end{equation}
\item[(2)] If $1\leq m\leq n$, calculate $C_m^{[k+1]}$ using equation 
\begin{equation}\label{08071}
c_m^{[k+1]}:=\bigg(\sum_{t=1}^T\Big(D_m\big(\varepsilon^{[k]}\big)\otimes\big(\tilde{B}_t^{[k]}+V_0^{[k]}\big)^{-1}\Big)\bigg)^{-1}\sum_{t=1}^T\Big(D_m(\varepsilon^{[k]}\big)\otimes \big(\tilde{B}_t^{[k]}+V_0^{[k]}\big)^{-1}\Big)J_m\tilde{\pi}_{0|t}^{[k]}
\end{equation}
and set $\bar{C}_{m}^{[k+1]}:=\big[C_{m}^{[k+1]}:0\big]$. Otherwise set $\bar{C}_m^{[k+1]}:=0$.
\item[(3)] For $i=1,\dots,n$ and $j=1,\dots,l$, calculate $\Big(\varepsilon_j^{[k+1]}\Big)^2$, $\Big(\alpha^{[k+1]}\Big)^2$, and $\Big(\gamma_i^{[k+1]}\Big)^2$ using equations 
\begin{eqnarray}\label{08072}
\Big(\varepsilon_j^{[k+1]}\Big)^2:=\frac{nT}{\displaystyle\sum_{t=1}^T\text{tr}\Big\{\Big(\tilde{\Lambda}_{0|t}^{[k]}\Delta_{\varepsilon_j}\otimes I_n\Big)+\Big(\nu_0^{[k]}+1\Big)\tilde{\Theta}_{|t}^{[k]}\Big(\Delta_{\varepsilon_j}\otimes \big(V_0^{[k]}+\tilde{B}_t^{[k]}\big)^{-1}\Big)\Big\}},
\end{eqnarray}
\begin{eqnarray}\label{08073}
\Big(\alpha^{[k+1]}\Big)^2:=\frac{n^2pT}{\displaystyle\sum_{t=1}^T\text{tr}\Big\{\Big(\tilde{\Lambda}_{0|t}^{[k]}\Delta_\alpha^{[k]}\otimes I_n\Big)+\Big(\nu_0^{[k]}+1\Big)\tilde{\Theta}_{|t}^{[k]}\Big(\Delta_\alpha^{[k]}\otimes \big(V_0^{[k]}+\tilde{B}_t^{[k]}\big)^{-1}\Big)\Big\}},
\end{eqnarray}
and
\begin{eqnarray}\label{08074}
\Big(\gamma_i^{[k+1]}\Big)^2:=\frac{npT}{\displaystyle\sum_{t=1}^T\text{tr}\Big\{\Big(\tilde{\Lambda}_{0|t}^{[k]}\Delta_{\gamma_i}^{[k]}\otimes I_n\Big)+\Big(\nu_0^{[k]}+T\Big)\tilde{\Theta}_{|t}^{[k]}\Big(\Delta_{\gamma_i}^{[k]}\otimes \big(V_0^{[k]}+\tilde{B}_t^{[k]}\big)^{-1}\Big)\Big\}}.
\end{eqnarray}
\item[(4)] For $p\geq 2$, to obtain $\beta^{[k+1]}$, solve the following nonlinear equation for $\beta$
\begin{eqnarray}\label{08075}
\sum_{\ell=1}^p\ln(\ell)=\frac{1}{nT}\sum_{t=1}^T\text{tr}\Big\{\Big(\tilde{\Lambda}_{0|t}^{[k]}\Delta_{\beta}^{[k]}\otimes I_n\Big)+\Big(\nu_0^{[k]}+1\Big)\tilde{\Theta}_{|t}^{[k]}\Big(\Delta_{\beta}^{[k]}\otimes \big(V_0^{[k]}+\tilde{B}_t^{[k]}\big)^{-1}\Big)\Big\}.
\end{eqnarray}
\item[(5)] To obtain $\nu_0^{[k+1]}$, solve nonlinear equation \eqref{08064} for $\nu_0$.
\item[(6)] Calculate $V_0^{[k+1]}$ using equation \eqref{08065}.
\item[(7)] Increase iteration count $k=k+1$ and go to step (2).
\end{itemize}
\end{algorithm}

\section{Conclusion}

Conditional matrix variate student $t$ distribution was introduced by \citeA{Battulga24g}. In this paper, we provide EM algorithms, which estimate parameters of the conditional matrix variate student $t$ distributions, including general case and special case with Minnesota prior. Also, we introduce a new conditional matrix variate student $t$ distribution, which is closely related to the \citeA{Battulga24g}'s conditional matrix variate student $t$ distribution. 

\section{Proofs of Results}

Here we provide proofs of Proposition \ref{prop02} and Lemma \ref{lem01}.

\begin{proof}[\textbf{Proof of Proposition \ref{prop02}}]
The proof follows \citeA{Battulga24g}. For given coefficient vector $\pi_t$, covariance matrix $\Sigma_t$, and information $\mathcal{F}_{t-1}$, conditional density functions of the random vector of endogenous variables $y_t$ is
\begin{eqnarray}\label{08076}
f(y_t|\pi_t,\Sigma_t,\mathcal{F}_{t-1})=\frac{|\Sigma_t|^{-1/2}}{(2\pi)^{n/2}}\exp\bigg\{-\frac{1}{2}\big(y_t-(\mathsf{Y}_t'\otimes I_{n})\pi_t\big)'\Sigma_t^{-1}\big(y_t-(\mathsf{Y}_t'\otimes I_{n}\big)\pi_t)\bigg\}.
\end{eqnarray}
Since conditional on $\mathcal{F}_0$, $(\pi_t,\Sigma_t)$ is independent of $\bar{y}_{t-1}$, conditional density functions of the random coefficient vector $\pi_t$ and covariance matrix $\Sigma_t$ are given by
\begin{eqnarray}\label{08077}
f(\pi_t|\Sigma_t,\mathcal{F}_{t-1})=\frac{|\Sigma_t|^{-d/2}}{(2\pi)^{nd/2}|\Lambda_{0}|^{n/2}}\exp\bigg\{-\frac{1}{2}\big(\pi_t-\pi_{0}\big)'\big(\Lambda_{0}^{-1}\otimes \Sigma_t^{-1}\big)\big(\pi_t-\pi_{0}\big)\bigg\}
\end{eqnarray}
and
\begin{equation}\label{08078}
f(\Sigma_t|\mathcal{F}_{t-1})=\frac{|V_{0}|^{\nu_{0}/2}}{\Gamma_{n}(\nu_{0}/2)2^{n\nu_{0}/2}}|\Sigma_t|^{-(\nu_{0}+n+1)/2}\exp\bigg\{-\frac{1}{2}\text{tr}\Big(V_{0}\Sigma_t^{-1}\Big)\bigg\}.
\end{equation} 
By the completing square method, a joint conditional density function of the random vectors $y_t$ and $\pi_t$ is
\begin{eqnarray}\label{08079}
f(y_t,\pi_t|\Sigma_t,\mathcal{F}_{t-1})&=& c_1|\Sigma_t|^{-(1+d)/2}\exp\bigg\{-\frac{1}{2}\Big(\pi_t-\tilde{\pi}_{0|t}\Big)'\big(\tilde{\Lambda}_{0|t}^{-1}\otimes \Sigma_t^{-1}\big)\Big(\pi_t-\tilde{\pi}_{0|t}\Big)\bigg\}\\
&\times&\exp\bigg\{-\frac{1}{2}\Big(y_t'\Sigma_t^{-1}y_t+\pi_{0}'\big(\Lambda_{0}^{-1}\otimes \Sigma_t^{-1}\big)\pi_{0}-\tilde{\pi}_{0|t}'\big(\tilde{\Lambda}_{0|t}^{-1}\otimes \Sigma_t^{-1}\big)\tilde{\pi}_{0|t}\Big)\bigg\},\nonumber
\end{eqnarray}
where $\tilde{\Lambda}_{0|t}^{-1}:=\mathsf{Y}_t\mathsf{Y}_t'+\Lambda_0^{-1}$ is a $(d\times d)$ matrix, $\tilde{\pi}_{0|t}:=\big((\tilde{\Lambda}_{0|t}\mathsf{Y}_t)\otimes \Sigma_t^{-1}\big)y_t+\big((\tilde{\Lambda}_{0|t}\Lambda_0^{-1})\otimes \Sigma_t^{-1}\big)\pi_0$ is an $([nd]\times 1)$ vector, and normalizing coefficient equals
\begin{equation}\label{08080}
c_1:=\frac{1}{(2\pi)^{n(1+d)/2}|\tilde{\Lambda}_{0|t}|^{n/2}}.
\end{equation}
If we integrate the above joint density function with respect to the vector $\pi_t$, then an integral, corresponding to the first exponential is proportional to $|\tilde{\Lambda}_{0|t}\otimes \Sigma_t|^{1/2}=|\mathsf{Y}_t\mathsf{Y}_t'+\Lambda_{0|t}^{-1})|^{n/2} |\Sigma_t|^{d/2}$. Therefore, we have that
\begin{eqnarray}\label{08081}
f(y_t|\Sigma_t,\mathcal{F}_{t-1})= c_2|\Sigma_t|^{-1/2}\exp\bigg\{-\frac{1}{2}\Big(y_t'\Sigma_t^{-1}y_t+\pi_{0}'\big(\Lambda_{0}^{-1}\otimes \Sigma_t^{-1}\big)\pi_{0}-\tilde{\pi}_{0|t}'\big(\tilde{\Lambda}_{0|t}^{-1}\otimes \Sigma_t^{-1}\big)\tilde{\pi}_{0|t}\Big)\bigg\},
\end{eqnarray}
where the normalizing coefficient equals
\begin{equation}\label{08082}
c_2:=\frac{1}{(2\pi)^{n/2}|\Lambda_{0}|^{n/2}|\tilde{\Lambda}_{0|t}^{-1}|^{n/2}}.
\end{equation}
Hence, according to the well--known formula that for suitable matrices $A,B,C,D$, 
\begin{equation}\label{08083}
\text{vec}(A)'(B\otimes C)\text{vec}(D)=\text{tr}(DB'A'C),
\end{equation}
we find that
\begin{equation}\label{08084}
f(y_t|\Sigma_t,\mathcal{F}_{t-1})=c_2|\Sigma_t|^{-1/2}\exp\bigg\{-\frac{1}{2}\text{tr}\big(\tilde{B}_t\Sigma_t^{-1}\big)\bigg\}.
\end{equation}
Thus, it follows from equations \eqref{08076} and \eqref{08078} that a joint conditional density of the random vector $y_t$ and random matrix $\Sigma_t$ is  
\begin{equation}\label{08085}
f(y_t,\Sigma_t|\mathcal{F}_{t-1})= c_3|\Sigma_t|^{-(\nu_0+n+2)/2}\exp\bigg\{-\frac{1}{2}\text{tr}\Big(\big(\tilde{B}_t+V_0\big)\Sigma_t^{-1}\Big)\bigg\},
\end{equation}
where the normalizing coefficient equals
\begin{equation}\label{08086}
c_3:=\frac{1}{(2\pi)^{n/2}}\frac{|V_0|^{\nu_0/2}}{|\Lambda_0|^{n/2}|\tilde{\Lambda}_{0|t}^{-1}|^{n/2}\Gamma_{n}(\nu_0/2)2^{n\nu_0/2}}.
\end{equation}
Consequently, a density function of the random vector $y_t$ is given by
\begin{eqnarray}\label{08087}
f(\bar{y}_t|\bar{s}_t,\mathcal{F}_0)=\int_{\Sigma_t>0}f(y_t,\Sigma_t|\mathcal{F}_{t-1})d\Sigma_t=c_3\prod_{k=1}^{r_t}\frac{\Gamma_{n}\big(\nu_{0t|t}/2\big)2^{n(\nu_0+1)/2}}{\big|\tilde{B}_t+V_0\big|^{(\nu_0+1)/2}}.
\end{eqnarray}
By the completing square method, the matrix $\tilde{B}_t$ can be written by 
\begin{eqnarray}\label{08088}
\tilde{B}_t&=&\big(y_t-\pi_0^\circ\Lambda_0^{-1}\tilde{\Lambda}_{0|t}\mathsf{Y}_t\phi_t^{-1}\big)\phi_t^{-1}\big(y_t-\pi_0^\circ\Lambda_0^{-1}\tilde{\Lambda}_{0|t}\mathsf{Y}_t\phi_t^{-1}\big)'\nonumber\\
&-&\pi_0^\circ\Lambda_0^{-1}\tilde{\Lambda}_{0|t}\mathsf{Y}_t\phi_t^{-1}(\mathsf{Y}_t)'\tilde{\Lambda}_{0|t}\Lambda_0^{-1}(\pi_0^\circ)'+\pi_0^\circ\Lambda_0^{-1}(\pi_0^\circ)'-\pi_0^\circ\Lambda_0^{-1}\tilde{\Lambda}_{0|t}\Lambda_0^{-1}(\pi_0^\circ)',
\end{eqnarray}
where $\phi_t:=1-\mathsf{Y}_t'\tilde{\Lambda}_{0|t}\mathsf{Y}_t$. We consider the following product
\begin{equation}\label{08089}
1_t:=\big(1+\mathsf{Y}_t'\Lambda_0\mathsf{Y}_t\big)\big(1-\mathsf{Y}_t'\tilde{\Lambda}_{0|t}\mathsf{Y}_t\big).
\end{equation}
It equals
\begin{eqnarray}\label{08090}
1_t&=&1+\mathsf{Y}_t'\Lambda_0\mathsf{Y}_t-\mathsf{Y}_t'\tilde{\Lambda}_{0|t}\mathsf{Y}_t-\mathsf{Y}_t'\Lambda_0\mathsf{Y}_t\mathsf{Y}_t'\tilde{\Lambda}_{0|t}\mathsf{Y}_t.
\end{eqnarray}
If we add and subtract the matrix $\Lambda_0^{-1}$ into the term $\mathsf{Y}_t\mathsf{Y}_t'$ in the last line of the above equation, then $1_t$ equals 1. Consequently, $1+\mathsf{Y}_t'\Lambda_0\mathsf{Y}_t$ is a reciprocal of $\phi_t$, that is,
\begin{equation}\label{08091}
\phi_t^{-1}=1+\mathsf{Y}_t'\Lambda_0\mathsf{Y}_t.
\end{equation}
Since it takes a positive value, the matrix $B_t$ is a positive semi--definite matrix. Now, we consider the term $\Lambda_0^{-1}\tilde{\Lambda}_{0|t}\mathsf{Y}_t\phi_t^{-1}$ in the first line in equation \eqref{08088}. Similarly as before, by adding and subtracting $\Lambda_0^{-1}$ into the term $\mathsf{Y}_t\mathsf{Y}_t'$, one obtains that
\begin{equation}\label{08092}
\Lambda_0^{-1}\tilde{\Lambda}_{0|t}\mathsf{Y}_t\phi_t^{-1}=\mathsf{Y}_t.
\end{equation}
Consequently, the second line of equation \eqref{08088} equals zero. Let $\Lambda_{0t}^{1/2}$ be the Cholesky factor of the matrix $\Lambda_0$, i.e., $\Lambda_0=\big(\Lambda_0^{1/2}\big)'\Lambda_0^{1/2}$. Then, according to the Sylvester's determinant theorem, see \citeA{Lutkepohl05}, $\phi_t^{-1}$ equals
\begin{equation}\label{08093}
|\phi_t^{-1}|=\big|1+\big(\Lambda_{0}^{1/2}\mathsf{Y}_t\big)'\Lambda_0^{1/2}\mathsf{Y}_t\big|=\big|I_d+\Lambda_0^{1/2}\mathsf{Y}_t\mathsf{Y}_t'\big(\Lambda_0^{1/2}\big)'\big|.
\end{equation}
That completes the proof of the Proposition.
\end{proof}

\begin{proof}[\textbf{Proof of Lemma \ref{lem01}}]
By the conditional probability formula and law of total probability, the density function $f(\pi_t,\Sigma_t|\mathcal{F}_T)$ is represented by
\begin{eqnarray}\label{08094}
f(\pi_t,\Sigma_t|\mathcal{F}_T)=\frac{\int_{\pi_{-t},\Sigma_{-t}}f(y,\pi,\Sigma|\mathcal{F}_0)d\pi_{-t}d\Sigma_{-t}}{f(y|\mathcal{F}_0)}
\end{eqnarray}
for $t=1,\dots,T$, where $\pi_{-t}$ is a $(d\times [T-1])$ matrix, which excludes the vector $\pi_t$ from a matrix $[\pi_1:\dots:\pi_T]$ and $\Sigma_{-t}$ is an $(n\times [(T-1)n])$ matrix, which excludes the matrix $\Sigma_t$ from a matrix $[\Sigma_1:\dots:\Sigma_T$. Due to the conditional probability formula and the assumption that for given initial information $\mathcal{F}_0$, $(\pi_1,\Sigma_1),\dots,(\pi_T,\Sigma_T)$ are independent, the numerator of the above equation equals
\begin{eqnarray}\label{08095}
&&\int_{\pi_{-t},\Sigma_{-t}}\prod_{i=1}^Tf(y_i|\pi_i,\Sigma_i,\mathcal{F}_{i-1})f(\pi_i,\Sigma_i|\mathcal{F}_0)d\pi_{-t}d\Sigma_{-t}\nonumber\\
&&=\prod_{i=1,i\neq t}^Tf(y_i|\mathcal{F}_{i-1})f(y_t|\pi_t,\Sigma_t,\mathcal{F}_{t-1})f(\pi_t,\Sigma_t|\mathcal{F}_0).
\end{eqnarray}
On the other hand, by the conditional probability formula, the denominator of equation \eqref{08094} equals $\prod_{i=1}^Tf(y_i|\mathcal{F}_{i-1})$. Consequently, since conditional on the initial information $\mathcal{F}_0$, $(\pi_t,\Sigma_t)$ is independent of the random vector $\bar{y}_{t-1}$, one obtains equation \eqref{08059}.
\end{proof}

\bibliographystyle{apacite}
\bibliography{References}

\end{document}